\documentstyle[sprocl,epsfig]{article}

\newcommand \beq{\begin{eqnarray}}
\newcommand \eeq{\end{eqnarray}}
\def\del{\partial}                              

\def\frac#1#2{{#1 \over #2}}
\def\simge{\mathrel{%
   \rlap{\raise 0.511ex \hbox{$>$}}{\lower 0.511ex \hbox{$\sim$}}}}
\def\simle{\mathrel{
   \rlap{\raise 0.511ex \hbox{$<$}}{\lower 0.511ex \hbox{$\sim$}}}}

\def\journal#1#2#3#4{\ {#1}{\bf #2}, {#4} ({#3})}
\def\NPB{\journal{Nucl.\ Phys.\ {\bf B}}}

\def\PLB{\journal{Phys.\ Lett.\ {\bf B}}}

\def\PRC{\journal{Phys.\ Rev.\ {\bf C}}}
\def\PRD{\journal{Phys.\ Rev.\ {\bf D}}}
\def\PRL{\journal{Phys.\ Rev.\ Lett.}}

\def\be{\begin{equation}}
\def\ee{\end{equation}}
\def\bea{\begin{eqnarray}}
\def\eea{\end{eqnarray}}

\begin{document}

\begin{titlepage}
\begin{flushright}
CERN-TH/99-83\\
hep-ph/9903486
\end{flushright}
\vspace*{1.5cm}
\begin{center}
\baselineskip=13pt
{\large{\bf {NON-PERTURBATIVE ASPECTS OF HOT QCD}}}
\vskip0.5cm
\vskip0.3cm
Edmond IANCU\footnote{E-mail: edmond.iancu@cern.ch}

{\it Theory Division, CERN\\ CH-1211, Geneva 23, 
Switzerland}

\end{center}

\vskip 1.8cm
\begin{abstract} 
I discuss some non-perturbative aspects of hot gauge theories
as related to the unscreened static magnetic interactions.
I first review some of the infrared divergences which cause
the breakdown of the perturbation theory. Then I show that
kinetic theory, as derived from quantum field theory, is a powerful
tool to construct effective theories for the soft modes,
which then can be treated non-perturbatively.
The effective theory at the scale $gT$ follows from a
collisionless kinetic equation, of the Vlasov type.
The effective theory at the scale $g^2T$ is generated by a
Boltzmann equation which includes the collision term for colour relaxation.
 \end{abstract}
\vskip 2.5cm

\begin{flushleft}
CERN-TH/99-83\\
March 1999\\
Invited talk at the Conference ``Strong and Electroweak Matter '98''
(Copenhagen, 2-5 Dec 1998).
\end{flushleft}
\end{titlepage}

\title{NON-PERTURBATIVE ASPECTS OF HOT QCD}

\author{Edmond Iancu}

\address{Theory Division, CERN\\ CH-1211, Geneva 23, 
Switzerland\\E-mail: Edmond.Iancu@cern.ch}

\maketitle\abstracts{
I discuss some non-perturbative aspects of hot gauge theories
as related to the unscreened static magnetic interactions.
I first review some of the infrared divergences which cause
the breakdown of the perturbation theory. Then I show that
kinetic theory, as derived from quantum field theory, is a powerful
tool to construct effective theories for the soft modes,
which then can be treated non-perturbatively.
The effective theory at the scale $gT$ follows from a
collisionless kinetic equation, of the Vlasov type.
The effective theory at the scale $g^2T$ is generated by a
Boltzmann equation which includes the collision term for
colour relaxation.}

\section{Introduction}

At  high temperature, the non-Abelian gauge theories describe
weakly coupled plasmas, which, in a first approximation,
are very much alike the ordinary, electromagnetic, plasmas
\cite{BIO96,MLB96}.
The plasma constituents, e.g., quarks and gluons for hot QCD, 
have typical momenta $k\sim T$, and take part in
collective excitations which typically develop on a
space-time scale  $\lambda\sim 1/gT$.
($T$ is the temperature,
and $g$ is the gauge coupling, assumed to be small.)
Such excitations are similar to the familiar
charge oscillations of the electromagnetic plasmas \cite{PhysKin}, and
can indeed be described by simple kinetic
equations of the Vlasov type \cite{qcd,BIO96}.

But this simple analogy breaks down at the softer scale $g^2 T$.
There, the non-Abelian plasmas enter
a new non-perturbative regime where the coupling constant is small
but the field strengths are large, so that perturbation theory breaks
down because of large non-linear effects. Indeed, gluons obey
Bose-Einstein statistics, so the population of the soft ($k\ll T)$ gluon
modes is strongly enhanced in thermal equilibrium:
\beq\label{BE}
N_0(k)\,\equiv\,\frac{1}{{\rm e}^{\beta k}-1}\,\simeq\,\frac{T}{k}
\,\,\gg \,1,\qquad {\rm for}\,\,\, k\ll T.\eeq
Thus, the long wavelength thermal fluctuations with $\lambda\sim 1/g^2 T$
involve many quanta $N_0\sim 1/g^2$ and behave in many respects
as classical colour fields $A^\mu_a$ with large amplitudes.
We shall verify later that, typically,
$|A| \equiv \sqrt{\langle A^2 \rangle }\sim gT$. This is
a large fluctuation in the sense that the two terms in the soft covariant
derivative $D_x\equiv \del_x + igA$ are of the same order in $g$, 
$\del_x \sim gA \sim g^2 T$, so that
the non-linear effects are indeed non-perturbative. 
In perturbation theory, such effects show up as infrared divergences
in relation with the mutual interactions of the soft ($k\sim g^2T$) magnetic
gluons (cf. Sec. 2.a below).

Moreover, infrared divergences are also associated with interactions
among the hard ($k\sim T$) particles, as mediated by the exchange of soft 
magnetic gluons (or photons, in QED). This occurs,
 for instance, in the calculation
of the quasiparticles damping rates in both QCD and QED (cf. Sec. 2.b).

In order to deal with such problems, one has to go beyond ordinary
perturbation theory. For an Abelian plasma, it is possible to
eliminate the infrared problem of the damping rate by a specific
resummation of the perturbation theory, based on the Bloch-Nordsieck (or eikonal)
approximation \cite{lifetime}. For non-Abelian plasmas, the non-linear
effects in the soft magnetic sector must be treated exactly, 
which requires numerical methods.
An useful strategy in this sense --- which is especially well suited
for real-time calculations ---  is to first construct an 
{\it effective theory} for the soft
fields, by integrating out the hard fields in perturbation theory,
and then study the effective theory non-perturbatively, e.g., 
as a classical theory on a lattice 
\cite{McLerran,AmbK,Yaffe,Hu,baryo,Bodeker98,ASY98,Moore98,Manuel99}. 
(The classical approximation will be discussed in Sec. 3 below.)
This strategy can be seen as an extension to 
real time of the ``dimensional reduction'' generally performed in static
calculations \cite{Kajantie}.

What I would like to show you here, is that kinetic theory is a powerful 
tool for constructing such effective theories. This has been
first demonstrated for the collective dynamics at the scale $gT$,
where we have shown  \cite{qcd} that simple,
collisionless, kinetic equations resum
an infinite number of one-loop diagrams with soft external lines
and hard loop momenta, the so-called ``hard thermal loops'' \cite{BP90,FT90}.
These equations will be reviewed below, in Sec. 4, together with the 
effective classical theory they generate \cite{baryo}, in Sec. 5.
Then, in Sec. 6, I shall extend this approach to describe
collective colour excitations at the softer scale $g^2T$. 
This involves a Boltzmann equation \cite{ASY98,BEQCD}
which generates  B\"odeker's effective theory \cite{Bodeker98} (see 
also Refs. \cite{Gyulassy93,Manuel99}).

\section{Infrared problems in perturbation theory}

\bigskip 
{\it (a) Higher-order corrections to the free energy}

\begin{figure}
\protect \epsfxsize=11.cm{\centerline{\epsfbox{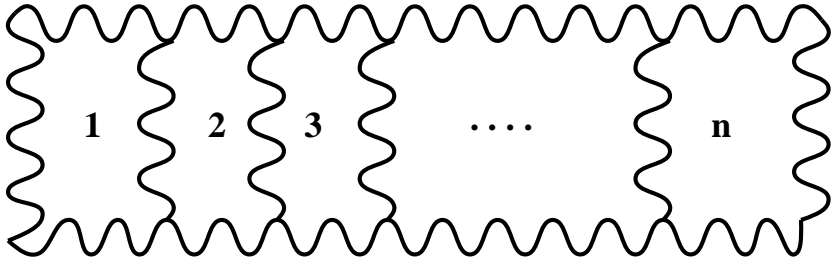}}}
         \caption{An $n$-loop ladder diagram contributing to the
free energy in hot QCD.}
\label{ladder}
\end{figure} 

\bigskip 
The infrared (IR) complications are most easily seen
in the calculation of static quantities, like the pressure,
as performed in the imaginary time formalism\footnote{
I consider a purely Yang-Mills plasma; indeed, quarks
are not important for the infrared physics to be discussed here.}.
There, the energies of the gluon modes are purely imaginary and discrete,
$k_0=i\omega_n\equiv i2\pi nT$, with integer $n$ (``Matsubara frequencies''),
which provides a large ``screening mass'' $|\omega_n| \sim T$ for all the
non-static ($n\ne 0$) modes. Thus, the IR problems can be associated only
with the static ($\omega_n=0$) Matsubara modes, and the most severe IR 
divergences are expected to come from diagrams involving the static modes alone.

For instance, when computing higher order corrections to the free energy
in hot QCD, one finds strong IR divergences from the ladder diagrams
depicted in Fig.~\ref{ladder}. In this diagram, all the propagators are static, 
and the loop integrations are three-dimensional. By power counting, this
can be estimated as
\cite{MLB96} $F^{(n)}\,\sim\,g^6T^4\left(g^2T/\mu\right)^{n-4}$
(for $n\ge 4$ loops), where $\mu$ is an ad-hoc IR cutoff.

For the {\it electric} gluons, an IR
cutoff $\sim gT$ is indeed generated dynamically, via Debye screening.
In the {\it magnetic} sector, however, screening can only occur
non-perturbatively, at the scale $g^2 T$.
With $\mu \sim g^2T$, all the ladder diagrams with four
or more loops will contribute to the {\it same} order in $g$, namely, 
to order $g^6$. Thus, perturbation theory breaks down, in the sense
that we lose the usual connection between
powers of the coupling constant and the number of loops.

As mentioned in the Introduction, this breakdown
is associated with long wavelength ($\lambda \sim 1/g^2 T$) thermal 
fluctuations with large amplitudes, $A\sim gT$. To see this, consider
the free propagator of the magnetic gluon, in imaginary time:
\beq
\langle A(\tau,{\bf x}) A(0)\rangle\,=\,
T\sum_n\int\frac{{\rm d}^3k}{(2\pi)^3}\,
{\rm e}^{-i\omega_n\tau +i{\bf k\cdot x}}\,\frac
{1}{k^2+\omega_n^2}\,.\eeq
By letting $\tau \to 0$, $x \to 0$, and keeping only the contribution
of the static modes ($\omega_n=0$) with soft momenta $k\sim g^2 T$,
one obtains:
\beq  \langle A^2 \rangle \,\simeq\,T\int\frac{{\rm d}^3k}{k^2}
\,\sim\,g^2T^2,\eeq
so that $|A|\equiv \sqrt{\langle A^2 \rangle }
\sim gT$, as anticipated.

A different perspective on these IR problems follows by
noticing that the diagram in Fig.~\ref{ladder} is actually a graph
of three-dimensional QCD with coupling constant $g_3=g\sqrt{T}$.
That is, the leading IR behaviour of hot QCD can be studied with
the replacement (with $A^i_a({\bf x})\equiv T\int_0^\beta {\rm d}\tau
A^i_a(\tau,{\bf x})$) :
\beq\label{4to3}
Z_4\,\equiv\,\int{\cal D}A^\mu_a(\tau,{\bf x})\,
\exp\left\{-\int_0^\beta {\rm d}\tau\int{\rm d}^3x \,\frac{1}{4}\,F^{\mu\nu}_a
F_{\mu\nu}^a\right\}\,\longrightarrow\,\nonumber\\
\longrightarrow\,\,Z_3\,\equiv\,\int{\cal D}A^i_a({\bf x})\,
\exp\left\{-\beta \int{\rm d}^3x \,\frac{1}{4}\,F^{ij}_a
F_{ij}^a\right\}\,\eeq
which reduces the initial finite-temperature problem in four dimensions
to an effective zero-temperature problem in three Euclidean dimensions.
This is the crudest example of ``dimensional reduction'', a strategy
which consists in integrating out the non-static modes in perturbation theory
in order to obtain an effective three-dimensional theory for
the static modes \cite{Kajantie}. This is convenient since
three-dimensional lattice simulations are much easier to perform than
in four dimensions. For instance,  the magnetic mass,
and the non-perturbative correction, of O$(g^6)$, to the pressure,
have been computed in this way \cite{Karsch}. Moreover, the reduction 
to three dimensions has also permitted for some analytic
non-perturbative studies of the magnetic screening \cite{KN96}.
\bigskip 

{\it (b) Quasiparticle damping rates}
\bigskip 

Non-perturbative aspects, as related to the unscreened magnetic gluons,
appear also in the computation of some {\it real-time} correlation functions.
The most celebrated example in that sense is the rate for anomalous baryon 
number violation at high temperature 
\cite{AmbK,Yaffe,Bodeker98,Moore98}, or ``hot sphaleron rate'',
a quantity which has been extensively discussed at this conference
(see, e.g., the contributions by P. Arnold, D. B\"odeker, A. Krasnitz,
G. Moore and L. Yaffe).

Here, however, I would like to discuss an example which is conceptually simpler
(since it involves only the computation of a 2-point function),
namely the calculation of the lifetime of the quasiparticles.
A single particle excitation is created by adding a particle with momentum
${\bf p}$ (e.g., a transverse gluon) to the plasma initially in equilibrium. 
The added particle will then scatter off the other particles in the
thermal bath (see Fig.~\ref{Born}), thus changing its momentum and colour.
That is, the initial excitation will decay, and this can be
measured from the corresponding retarded propagator $D_R(t,{\bf p})
\equiv i \theta(t)\langle [A(t,{\bf p}),A(0,-{\bf p})]\rangle$.
A usual expectation is that $D_R(t,{\bf p})$ decays exponentially in 
time, $|D_R(t,{\bf p})|^2 \sim {\rm e}^{-2\gamma t}$,
which identifies the lifetime of the single particle excitation as
$\tau =1/2\gamma$. It turns out, however, that the calculation of $\gamma$
is afflicted with IR problems. Specifically \cite{BP90,lifetime} :
\begin{figure}
\protect \epsfysize=3.cm{\centerline{\epsfbox{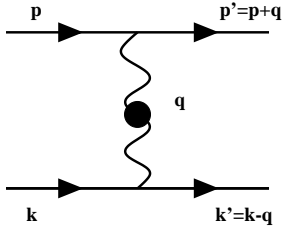}}}
         \caption{Elastic scattering in the (resummed) Born approximation.
The continuous lines refer to hard gluons, while the wavy line is the soft gluon
exchanged in the collision. The blob indicates the resummation of the screening
effects.}
\label{Born}
\end{figure}
\beq\label{G2L}
\gamma \simeq\, \frac{g^4N_c^2 T^3}{12}\,
\int{\rm d}q  \int_{-q}^q\frac{{\rm d}q_0}{2\pi}
\left\{|{}^*{\cal D}_l(q_0,q)|^2\,+\,\frac{1}{2}\left(1-\frac{q_0^2}{q^2}\right)^2
|{}^*{\cal D}_t(q_0,q)|^2\right\},\eeq
where ${}^*{\cal D}_{l}(q)$ and ${}^*{\cal D}_{t}(q)$ denote the 
propagators of the exchanged gluon
in the electric and the magnetic channels, respectively.
These are dressed so as to include the
screening effects at the scale $gT$; indeed, with a bare gluon
propagator, the integral in eq.~(\ref{G2L}) would be quadratically
IR divergent, showing that the damping rate is dominated by
soft momentum transfers, $q\simle gT$. The dressed propagators
have the following IR behaviour \cite{BIO96,MLB96}
(below, $m_D$ is the Debye mass, $m_D^2=g^2 N_c T^2/3$):
 \beq\label{DSTAT}
{}^*{\cal D}_l(q_0\to 0,q)\,\simeq\,\frac{- 1}{q^2 + m_{D}^2}\,,
\qquad {}^*{\cal D}_t(q_0\ll q)\simeq\,\frac{1}
{q^2-i\,(\pi q_0/4q)\,m_D^2}\,,\eeq
which exhibits Debye screening in the electric sector
and dynamical screening in the magnetic sector. The latter is due
to Landau damping \cite{PhysKin}, i.e., the thermal absorbtion
of the space-like gluons, which gives an imaginary part to the 
self-energy of the magnetic gluon: $
{\rm Im}\,\Pi_t(q_0\ll q) \simeq -(\pi q_0/4q)\,m_D^2$.
This is proportional to the frequency $q_0$ since only the time-dependent
($q_0\ne 0$) magnetic fields can transfer energy to the plasma constituents
(as mechanical work), and thus get damped.

Because of Debye screening, the electric contribution to the damping rate
$\gamma_l$ is finite and of order $g^4 T^3/m_D^2 = {\rm O}(g^2T)$.
However, even after including the dynamical screening, the magnetic
contribution $\gamma_t$ remains logarithmically divergent (below,
$\mu$ is an ad-hoc IR cutoff) :
\beq\label{G2LR}
\gamma_t \,\simeq\,
\frac{g^2N_cT}{4\pi}\,\int_{\mu}^{m_D}\frac{{\rm d}q}{q}\,=\,
\frac{g^2 N_c T}{4\pi}\,\ln\frac{m_D}{\mu}\,.\eeq
The remaining divergence
is associated to the static magnetic interactions, which are not
screened at the scale $gT$. Assuming magnetic screening at the
scale $g^2 T$, it follows that $\gamma \simeq \alpha N_cT\ln(1/g)$
(with $\alpha\equiv g^2/4\pi$), where, however, the constant
term under the logarithm cannot be determined (since sensitive
to the non-perturbative screening mechanism).
Thus, in QCD, the lifetime of the quasiparticles cannot be computed
in perturbation theory beyond logarithmic accuracy.

The same IR problem occurs also in QED, in the calculation
of the lifetime of the charged particles. There, this is even more
intriguing, since there is no magnetic screening in the Abelian 
theories \cite{BIP95}. However, as shown in Ref. \cite{lifetime},
the Abelian problem can be solved by a further resummation of the
perturbation theory, with the peculiar result that the electron propagator
has an anomalous, non-exponential, decay law:  $G_R(t,{\bf p})
\sim {\rm exp}\{- \alpha Tt\ln(m_D t)\}$ (see also Ref. \cite{BDV98}).

\section{The classical approximation}

I come now to the main question to be addressed in this talk,
which is, how to compute non-perturbative {\it real-time} 
correlations in hot gauge theories. Clearly, the standard
lattice calculations, as formulated in imaginary time,
are not appropriate for this problem. Fortunately, a fully quantum
calculation is actually not needed: because of the Bose enhancement,
the non-perturbative modes with $k\sim g^2 T$ have
large thermal occupation numbers (cf. eq.~(\ref{BE})), which, by the 
correspondence principle, is a classical limit.
For instance, the average energy per soft mode 
in thermal equilibrium, namely,
\beq\label{EQP}
\varepsilon(k)\,=\,\frac{k}{{\rm e}^{\beta k}-1}\,\simeq\,\,{T}
\qquad{\rm for}\,\,\,\, \,\,k\ll T.
\eeq
is the same as expected from the classical equipartition theorem.
Based on this observation, it has been suggested to compute
the hot baryon number violation rate through lattice simulations
of a {\it classical} thermal field theory \cite{McLerran,AmbK}.
The only question is, what is the correct classical theory ?

It is well known that the classical approximation becomes 
meaningless at high momenta $k\simge T$, where eq.~(\ref{BE}) is not
correct anylonger. At a first sight, one could 
expect this to be irrelevant for the problem at hand. Indeed, we are
interested here in non-perturbative correlations, as determined by the
dynamics at the scale $g^2 T$. Thus, one may expect such correlations
not to be sensitive to the hard plasma modes \cite{AmbK}.
If this was the case, such quantities could be simply
computed from the classical Yang-Mills theory at finite temperature,
without worrying too much about its bad ultraviolet behaviour.

But the previous examples in Sec. 2 show
that this argument is
too na\"{\i}ve \cite{Yaffe} : in the plasma, the soft and hard modes are
coupled by the interactions, which results
in screening effects which considerably modify
the dynamics of the soft modes. For instance,
eq.~(\ref{DSTAT}) for ${}^*{\cal D}_t(q)$ shows that, for large enough
frequencies $q_0$, the soft magnetic fields are efficiently screened by 
Landau damping, and therefore decouple from 
the non-perturbative IR  physics.
For $q\sim g^2 T$, only the modes with very low frequencies
$q_0 \simle q^3/m_D^2 \sim g^4 T$ can take part in non-perturbative
phenomena. According to Arnold, Son and Yaffe \cite{Yaffe}, this sets
the time scale for non-perturbative phenomena to be $1/g^4 T$
(see also Refs. \cite{lifetime,Bodeker98,ASY98}).

The classical Yang-Mills theory does not describe correctly the
screening effects due to the hard particles. For instance, it yields
a linearly divergent Debye mass \cite{McLerran,Yaffe,baryo}
$m_{cl}^2 \sim g^2 T\Lambda$, rather than the correct, quantum,
result  $m_D^2 \sim g^2 T^2$.
A possible solution to this problem is to treat hard and soft modes
on a different footing: first, the hard modes are integrated out
in perturbation theory, which properly generates the screening corrections;
then, the resulting effective theory for the soft modes is treated as
a classical field theory, via non-perturbative methods (e.g., via lattice
simulations). This strategy requires an unambiguous separation
between hard and soft degrees of freedom, e.g., an intermediate cutoff
$\mu$, which moreover must be consistent with gauge symmetry and with
the lattice implementation. If $\mu$ is chosen such as $gT \ll \mu \ll T$, 
then one obtains the ``hard thermal loop'' (HTL) effective theory 
\cite{BP90,FT90,qcd,McLerran,baryo}, to be presented in the next two sections.
If, on the other hand, $g^2T \ll \mu \ll gT$, then one obtains B\"odeker's
effective theory \cite{Bodeker98,ASY98,Manuel99,BEQCD}, to be
discussed in Sec. 6 below.

\section{Effective theories from kinetic equations}

To construct an effective theory for the soft modes,
one needs to study the dynamics of the hard plasma constituents
--- here, transverse gluons with momenta $k\sim T$ ---
 in the presence of soft ($q\simle gT$) background
fields $A^\mu_a$. This is an off-equilibrium situation:
the plasma is perturbed away from equilibrium by the background
fields which induce long wavelength ($\lambda \simge 1/gT$)
fluctuations in the colour density of the hard particles.
Since $\lambda \gg \bar r$ (where $\bar r \sim 1/T$ is the mean
interparticle distance), these are {\it collective} colour
excitations. Since, furthermore, $\lambda \gg \lambda_T$
(where $\lambda_T \equiv 1/k \sim 1/T$ is the thermal wavelength
of the hard particles), we expect such excitations to be described
by kinetic theory \cite{BIO96,PhysKin}. And, indeed, the equations
to be presented below \cite{qcd} can be viewed as a generalization
of the Vlasov equation for ordinary plasmas \cite{PhysKin}.

Specifically, the longwavelength  colour 
excitations of the hard transverse gluons
are described by a colour density matrix
$N_{ab}({\bf k},x)$ which, to the order of interest,
can be written in the form:
\beq\label{dn0}
N_{ab}({\bf k}, x)\,=\,N_0(\varepsilon_k)\delta_{ab}
 - gW_{ab}(x,{\bf v})\,({\rm d}N_0/{\rm d}
\varepsilon_k),\eeq
where $N_0(\varepsilon_k)\equiv 1/({\rm e}^{\beta \varepsilon_k}-1)$
is the equilibrium distribution (with $\varepsilon_k
=|{\bf k}|$), and the function 
$W(x,{\bf v})$, which parametrizes the off-equilibrium deviation,
is a colour matrix in the adjoint representation,
$W(x,{\bf v})\equiv W_a(x,{\bf v}) T^a$, which depends
upon the velocity ${\bf v}={\bf k}/\varepsilon_k$
(a unit vector), but not upon the
magnitude $k\equiv |{\bf k}|$ of the  momentum. It satisfies
the following simple equation:
\beq\label{W}
(v\cdot D_x)_{ab}W^b(x,{\bf v})\,=\,{\bf v}\cdot{\bf E}_a(x),\eeq
where $v^\mu\equiv (1,\,{\bf v})$, 
$D^\mu=\del^\mu+igA^\mu_a T_a$ is the covariant derivative defined
by the background field, and $E_a^i$ is the chromoelectric mean field.
The system is closed by the  Yang-Mills equations for the soft fields
$A_a^\mu$, namely:
\beq\label{ava0}
(D_\nu F^{\nu\mu})_a(x)&=&j^\mu_a(x),\eeq
with the induced current:
\beq\label{j}
j_a^\mu(x)&=&m_D^2\int\frac{{\rm d}\Omega}{4\pi}
\,v^\mu\,W_a(x,{\bf v}),\eeq
where the angular integral $\int {\rm d}\Omega$ runs over the 
orientations of ${\bf v}$. 

Eq.~(\ref{W}) is a collisionless kinetic equation. It has been obtained \cite{qcd}
from the general Dyson-Schwinger equations for the off-equilibrium
plasma, by neglecting the collisions among the hard particles
and by performing a gauge-covariant gradient expansion which takes profit
of the assumed separation of scales. Note that all these approximations
are controlled by the same small parameter, the coupling strength $g$,
so that eq.~(\ref{W}) is actually correct to leading order in $g$.

By formally solving eq.~(\ref{W}), 
we can express the current in terms of the gauge fields $A_a^\mu$,
and thus obtain an {\it effective} Yang-Mills equation which involves 
the soft fields alone:
\beq\label{ava}
D_\nu F^{\nu\mu}\,=\,m_D^2\int\frac{{\rm d}\Omega}{4\pi}\,
\frac{v^\mu v^i}{v\cdot D}\,E^i.\eeq
Eq.~(\ref{ava}) describes the propagation of soft colour fields
in the high-$T$ plasma. The hard particles are not explicit anymore, 
since they have been integrated to yield the induced current in the r.h.s.
By expanding this current in powers of the gauge fields 
one generates \cite{qcd} all the HTL's of Braaten and Pisarski \cite{BP90},
which encompass the screening phenomena at the scale $gT$ (cf. Sec. 2).
However, because of the  non-local structure of the current
(note the covariant derivative in the denominator), eq.~(\ref{ava}) is
not very convenient for the construction
of the classical thermal theory, to which I turn now.

\section{The classical  effective theory}

I shall now use eqs.~(\ref{W})--(\ref{j}) to define a classical
field theory at finite temperature \cite{baryo}. 
As usual with gauge theories, this is most
easily done in the temporal gauge $A^a_0=0$, where the 
equations read:
\beq\label{CAN}
E^a_i&=&-\del_0 A^a_i,\nonumber\\
-\del_0 E^a_i +\epsilon_{ijk}(D_j B_k)^a &=&
m_D^2\int\frac{{\rm d}\Omega}{4\pi}\,v_i\,W^a(x,{\bf v}),\nonumber\\
\left(\del_0 + {\bf v\cdot D}\right)^{ab} W_b&=&{\bf v \cdot E}^a,\eeq
together with the constraint expressing Gauss' law (i.e., the $\mu=0$
component of eq.~(\ref{ava0})) :
\beq\label{GAUSS}
G^a({\bf x})\equiv
({\bf D\cdot E})^a\,+\,m_D^2\int\frac{{\rm d}\Omega}{4\pi}\,W^a(x,{\bf v})
\,=\,0.
\eeq
Note that eqs.~(\ref{CAN}) are not in canonical form: this is already
obvious from the fact that we have an odd number of equations.
Still, it can be verified that these equations are conservative;
the corresponding, conserved energy functional has the 
gauge-invariant expression \cite{qcd,BIO96} :
\beq\label{H}
H=\frac{1}{2}\int {\rm d}^3 x\biggl\{{\bf E}_a\cdot{\bf E}_a\,+\,
{\bf B}_a\cdot{\bf B}_a\,+\,m_D^2
\int\frac{{\rm d}\Omega}{4\pi}\,W_a(x, {\bf v})\,W_a(x, {\bf v})\biggr\}.\eeq
Eqs.~(\ref{CAN})--(\ref{H}) define an effective theory for
the soft degrees of freedom. Besides the soft colour fields $A^i_a(x)$ and 
$E^i_a(x)$, these equations also involve the auxiliary fields $W_a(x, {\bf v})$
which simulate the hard thermal gluons (or, more precisely,
their long wavelength colour fluctuations). We are interested
in computing the (non-perturbative) real-time  correlations of the
fields $A^i_a$. To this aim, we need to construct the 
thermal partition function for this classical field theory.

Recall first how this
is done in some generic theory: the thermal expectation values
are obtained by first solving the classical equations of motion for given 
initial conditions, and then averaging over the initial conditions
(i.e., over the classical ``phase-space'')
with the Boltzmann weight exp$(-\beta H)$.
Since the initial conditions (say $\phi({\bf x})$ 
and $\dot\phi({\bf x})$ for a scalar theory)
depend only on the spatial coordinate ${\bf x}$, the phase space integration
is actually a {\it three-dimensional} functional integral, which can be
implemented on a lattice in the standard way.

For the problem at hand, the classical phase-space
is determined by the initial conditions to eqs.~(\ref{CAN}), that is,
\beq\label{INIT}
A_a^i(0,{\bf x})={\cal A}^i_a({\bf x}),\qquad
E_a^i(0,{\bf x})={\cal E}_a^i({\bf x}),\qquad
W_a(0,{\bf x,v})={\cal W}_a({\bf x,v}),\eeq
and the thermal weight is provided by the Hamiltonian in eq.~(\ref{H}).
Then,  the  (real-time) thermal correlation functions of the fields
$A^i_a$ can be obtained from the following generating functional:
\beq\label{Z}
Z_{cl}[J^a_i]\,=\,
\int {\cal D}{\cal E}^a_i\,{\cal D}{\cal A}^a_i\,{\cal D}{\cal W}^a\,
\delta({\cal G}^a)\,
\exp\left\{-\beta{\cal H}\,+\,\int{\rm d}^4x J^a_i(x) A^a_i(x)\right\},\eeq
where $A^i_a(t,{\bf x})$ is the solution to eqs.~(\ref{CAN})
with the initial conditions (\ref{INIT}), and ${\cal G}^a$ and ${\cal H}$
are expressed in terms of the initial fields, cf.
eqs.~(\ref{GAUSS}) and (\ref{H}).
Physically, the fluctuations in the initial 
conditions ${\cal W}({\bf x},{\bf v})$
for the auxiliary fields can be interpreted as a thermal noise due to the
hard particles and which drives the soft fields toward thermal equilbrium.
It is this noise which generates the Landau damping
of the soft correlation functions \cite{baryo}.

In particular, for $J=0$, eq.~(\ref{Z}) yields the following 
expression for the free energy
of the classical thermal radiation:
\beq\label{ZRED}
Z_{cl}=\int {\cal D}{\cal A}^a_0 {\cal D}{\cal A}^a_i
\exp\left\{-\beta\int{\rm d}^3x \left(\frac{1}{4}
({\cal F}^a_{ij})^2 + \frac{1}{2}({\cal D}_i {\cal A}_0)^2
+\frac{m_{D}^2}{2} ({\cal A}_0^a)^2 \right)\right\},\,\,\,\,\eeq
which is also the result expected from dimensional reduction
(see eq.~(\ref{4to3})).

To complete the construction of the effective theory, eq.~(\ref{Z})
must be supplemented with an UV cutoff $\mu$, which is the scale
separating hard from soft degrees of freedom: $gT\ll \mu \ll T$.
Correspondingly, the Hamiltonian must be extended to include 
$\mu$-dependent counterterms, chosen so as to cancel the cutoff
dependence of the classical theory in any complete calculation.
In practice, however, this ``matching'' turns out to be difficult
to achieve \cite{baryo,Yaffe}, mainly because of the constraints
of the lattice implementation. It is therefore rewarding that the 
results of the first lattice simulations \cite{Kari99}
of eqs.~(\ref{CAN})--(\ref{Z})
appear to be quite robust and insensitive to lattice artifacts.
In particular, the hot sphaleron rate obtained in this way
is consistent with the previous calculations in Ref. \cite{Hu}.

\section{A Boltzmann equation for colour}

Since the ``semi-hard'' modes with $q\sim gT$ are also perturbative,
it is possible to integrate them out as well, and thus get
an effective theory involving only the ``ultrasoft'' modes with 
$q\sim g^2 T$  \cite{Bodeker98}.
Then, the collisions among the hard particles, 
as mediated by the semi-hard fields, must be included explicitely in the 
kinetic equation. That is, the previous equation (\ref{W}) must 
be generalized so as to include the collisions terms.
This is also necessary for consistency: for colour fluctuations
at the scale $g^2 T$, the effects of the collisions among the plasma
particles become as important as those  of the mean fields \cite{BEQCD}.

The resulting kinetic equation is a Boltzmann equation describing
the propagation and relaxation of longwavelength
($\lambda \sim 1/g^2 T$) colour excitations.
It allows, in particular, to compute the colour conductivity
\cite{Bodeker98,ASY98,BEQCD}, thus clarifying some previous work on
this subject \cite{Gyulassy93}.
To leading logarithmic accuracy (see below), this equation has been
first obtained by B\"odeker \cite{Bodeker98}. Then, Arnold, Son and Yaffe
have shown \cite{ASY98} that B\"odeker's theory can be generated by a
rather simple Boltzmann equation, which has been physically motivated,
but not rigorously proven, in Refs. \cite{ASY98}. 
Recently, we have given a derivation
of this equation \cite{BEQCD}, starting from the quantum field equations.
This has clarified the nature of the approximations involved, thus fixing
its range of applicability. 

Remarkably, even after the inclusion of the collision term,
the density matrix $N_{ab}({\bf k},x)$ preserves the same structure
as in the mean field approximation (cf. eq.~(\ref{dn0})), but the
functions $W_a(x,{\bf v})$ satisfy a more complicated equation
(compare to eq.~(\ref{W})) :
\beq\label{W10}
(v\cdot D_x)^{ab}W_b(x,{\bf v})&=&{\bf v}\cdot{\bf E}^a(x)-
\gamma\left\{W^a(x,{\bf v})-\frac{\left\langle
\Phi({\bf v\cdot v}^\prime)W^a(x,{\bf v}^\prime)\right\rangle}
{\left\langle\Phi({\bf v\cdot v}^\prime)\right\rangle}\right\}.\eeq
The new feature here
is the collision term in the r.h.s. This
is proportional to the quasiparticle damping
rate $\gamma$ (cf. Sec. 2.b), which thus appears to set 
the scale for the colour relaxation time: $\tau_{col}\sim 1/\gamma \sim
1/(g^2T\ln(1/g))$. The other notations above are as follows:
the angular brackets in the collision term denote angular average
over the directions of the unit vector ${\bf v}^\prime$ (as 
in eq.~(\ref{gamma110}) below), and the
quantity $\Phi({\bf v\cdot v}^\prime)$ is given by:
\beq\label{PHII}\Phi({\bf v\cdot v}^\prime)\equiv
\int\frac{{\rm d}^4 q}{(2\pi)^2}
\delta(q_0- {\bf q\cdot v})
\delta(q_0- {\bf q\cdot v}^\prime)
\Big|{}^*{\cal D}_l(q)+ ({\bf v}_t\cdot{\bf v}_t^\prime)
\,{}^*{\cal D}_t(q)\Big|^2.\,\,\,\eeq
with ${}^*{\cal D}_{l}(q)$ and ${}^*{\cal D}_{t}(q)$ 
defined after eq.~(\ref{G2L}).
Up to a normalization, $\Phi({\bf v\cdot v}^\prime)$ is the total interaction
rate for two hard particles with momenta ${\bf k}$ and ${\bf p}$ (and
velocities ${\bf v}\equiv \hat{\bf k}$ and
${\bf v}^\prime \equiv \hat{\bf p}$) in the (resummed) Born approximation,
as illustrated in Fig.~\ref{Born}
(${\bf v}_t$ and ${\bf v}_t^\prime$ are the transverse projections
of the velocities with respect to the momentum ${\bf q}$ of the
exchanged gluon). The damping rate $\gamma$ is obtained
from $\Phi({\bf v\cdot v}^\prime)$ as follows (cf. eq.~(\ref{G2LR})) :
\beq\label{gamma110}
\gamma\,=\,\frac{g^4 N_c^2 T^3}{6}\int\frac{{\rm d}\Omega'}{4\pi}
\,\Phi({\bf v\cdot v}^\prime)\,\simeq\,\frac{g^2 N_c T}{4\pi}
\ln(1/g).\eeq
There are two important remarks about the previous equations:

First, whereas most transport
phenomena are dominated by large momentum transfers
$gT \simle q \simle T$, so that the typical
relaxation times are \cite{Baym90} $\tau_{tr} \sim 1/(g^4 T\ln(1/g))$,
the relaxation of colour excitations turns out to be
dominated by soft gluon exchanges, $g^2 T \simle q \simle gT$,
as the quasiparticle damping rate. There is a simple physical reason for that
\cite{Gyulassy93,Bodeker98,ASY98} : unlike momentum fluctuations,
which require a large angle scattering to relax, colour can be
efficiently exchanged in any scattering, even a small angle one.
This yields a colour conductivity \cite{Gyulassy93,Bodeker98,ASY98}
$\sigma_c \sim T/\ln(1/g)$, to be contrasted with the usual, electric,
conductivity \cite{Baym90} : $\sigma_{el} \sim T/(e^2\ln(1/e))$.
Note also that it is the same physical process
--- namely, the scattering via one-gluon exchange in Fig.~\ref{Born}
--- which provides relaxation for both single-particle and
collective (momentum or colour) excitations. If, nevertheless,
the relevant time scales turn out not to be the same
(namely $\tau \sim \tau_{col} \gg \tau_{tr}$),
it is because of specific cancellations among various collision terms \cite{BEQCD}, 
which occur in the calculation of most transport coefficients, but not 
in that of the quasiparticle lifetime $\tau$, 
or in that of the relaxation time of colour excitations
$\tau_{col}$.

Second, strictly speaking, the functional form of the collision
term in eq.~(\ref{W10}) is only valid to leading logarithmic accuracy,
because of the approximations performed in its derivation.
This limitation comes from the poor convergence of the  gradient
expansion when the range of the interactions becomes comparable 
to the scale of the system inhomogeneities \cite{BEQCD}.
Within this logarithmic accuracy, eq.~(\ref{W10}) can be shown
\cite{ASY98,BEQCD} to reduce to B\"odeker's equation \cite{Bodeker98}.

Note finally that, even though conceptually interesting,
the leading logarithmic approximation appears to be of little use
for the calculation of the hot baryon number violation rate.
Indeed, the numerical calculations
\cite{Hu,Moore98,Kari99} show that, for realistic values of $g$,
the constant term under the logarithm is sensibly larger than $\ln(1/g)$.

\end{document}